\documentclass[10.75pt]{article}

\usepackage{amsfonts}
\usepackage{amssymb}
\usepackage{amsbsy} 
\usepackage{bm}
\usepackage{rotating}
\usepackage{setspace}
\usepackage{citesort}
\usepackage{cites}
\usepackage{subscript}
\usepackage{subfigure}
\usepackage{float}
\usepackage{algorithmic}
\usepackage[centertags]{amsmath}
\usepackage{amsthm}
\usepackage[mathcal]{euscript}
\usepackage{booktabs}
\usepackage{tabls}
\usepackage{multirow}
\usepackage{tabularx}
\usepackage{threeparttable}
\usepackage[hang]{caption}
\usepackage[subnum]{cases}
\usepackage{graphicx}
\usepackage[margin=1in]{geometry}

\renewcommand{\refname}{References}

\newcommand{\bA}[1]{\ensuremath{\mathbf{#1}}}

\newcommand{\on}{\ensuremath{\hat{\Omega}}}
\newcommand{\onp}{\ensuremath{{\hat{\Omega}}}^{\prime}}

\newcommand{\Sn}{\ensuremath{\text{S}_{N}}~}

\newcommand{\Pl}{\ensuremath{\text{P}_{L}}~}

\newcommand{\nfp}{\ensuremath{\nabla^{2}}}

\newcommand{\lfp}{\ensuremath{\mathcal{L}_{\text{FP}}}}

\newcommand{\LEQ}[1]{\label{eq:#1}}
\newcommand{\EQ}[1]{(\ref{eq:#1})}
\newcommand{\REQ}[1]{\ref{eq:#1}}

\usepackage{float}
\floatstyle{ruled}
\newfloat{algo}{H}{loa}
\floatname{algo}{Algorithm}

\makeatletter
\renewcommand{\@thesubfigure}{\thefigure\thesubfigure\space}
\makeatother

\begin{document}

\bibnum{p}

\onehalfspacing

\thispagestyle{empty}

\begin{center}

\textbf{\large{Acceleration of the \Sn Equations with Highly Anisotropic Scattering using
the Fokker-Planck Equation}}

\vspace{2ex}

\begin{small}

$\mathrm{Japan~K.~Patel,}^1 ~\mathrm{James~S.~Warsa,}^2 ~\mathrm{and~ Anil~K.~Prinja}^3$

\vspace{1ex}

$^1 ~\mathrm{Department~of~Mechanical~and~Aerospace~Engineering,~The~Ohio~State~University,~Columbus,~OH}$ \\
$^2 ~\mathrm{Transport~Methods~Group~(CCS-2),~Los~Alamos~National~Laboratory, ~Los~Alalmos,~NM}$ \\
$^3 ~\mathrm{Department~of~Nuclear~Engineering,~The~University~of~New~Mexico,~Albuquerque,~NM}$ \\

\vspace{1ex}
patel.3545@osu.edu, warsa@lanl.gov, prinja@unm.edu
% \today\\
\end{small}

\vspace{2ex}
\end{center}

% ABSTRACT 

\textbf{Abstract}: The discrete ordinates method can model forward-peaked transport problems accurately. However, convergence of discrete ordinates solution can become arbitrarily slow upon use of standard iterative procedures like source iteration and GMRES. Standard zeroth and first moment-based acceleration methods like nonlinear diffusion acceleration and diffusion synthetic acceleration are ineffective in accelerating such problems because these methods do not correct higher order Legendre-moments of angular flux. We explore the idea of using Fokker-Planck as a preconditioner to accelerate forward-peaked transport problems in this paper. \\


%============================================================================
%
% Introduction
%
% +++++++++++++++++++++++++++++++++++++++++++++++++++++++++++++++++++++++++++++

\section{Introduction}

Transport problems with forward-peaked scattering kernels are encountered in several applications related to plasma physics, radiation sheilding, medical physics, and astrophysics. Such problems have extremely small mean free paths and nearly singular differential scattering cross-sections in the forward direction. Use of discrete ordinates method with standard methods like source iteration and GMRES can become extremely inefficient due to these properties. Standard acceleration techniques like diffusion synthetic acceleration (DSA) (Alcouffe, 1977) and nonlinear diffusion acceleration (NDA) (Smith et al., 2011) are ineffective in accelerating such problems because they assume all moments higher than the zeroth moment are inconsequential to the convergence of the solution. We see, later in this paper, that such an assumption becomes invalid for forward-peaked problems. \\

Several innovations have been made to accelerate the convergence of such problems. Valougeorgis, Williams, and Larsen (Valougeorgis et al., 1988) presented their work on stability analysis of \Pl acceleration applied to anisotropic neutron transport problems. This paper presented an extremely valuable framework for theoretical development and testing of future acceleration methods for transport problems with anisotropic scattering. Khattab and Larsen presented their modified \Pl acceleration method that used a modified form of \Pl equations with modified scattering cross section moments in (Khattab and Larsen, 1991). Morel and Manteuffel presented their angular multigrid method for solution of problems with high anisotropy in (Morel and Manteuffel, 1991). The angular multigrid method proved to be effective in 1D with a maximum spectral radius of $0.6$. The method however had to be modified later to preserve stability for problems in higher spatial dimensions in (Pautz et al., 1998). Turcksin and Morel integrated diffusion synthetic acceleration and angular multigrid to develop their diffusion synthetic acceleration-angular multigrid method in (Trucksin, 2012).  \\

Several approximations to the transport equation have also been derived to tackle forward-peaked problems. Most prominent of these is the Fokker-Planck approximation which is an asymptotic limit of the Boltzmann equation (Pomraning, 1992) as scattering angle and energy loss become diminishingly small (Fokker-Planck limit). Renormalization techniques can be applied for generating stable higher order approximations in this limit to obtain generalized-Fokker-Planck equations (Pomraning, 1996), (Prinja and Pomraning, 2001), and (Leakes and Larsen, 2001). The scattering kernel can be decomposed into smooth and singular parts (Caro and Ligou, 1983), (Landesman and Morel, 1989), (Aristova and Gol'din, 1998), (Dixon, 2015) to derive the Boltzmann-Fokker-Planck or Boltzmann-Fokker-Planck-like approximations. \\

In this paper, we primarily focus on the acceleration side of the solving forward-peaked problems. Problems with forward-peaked scattering kernels require acceleration of all slowly-converging Legendre-moments of angular flux with significant magnitudes. We  develop and test a synthetic acceleration method - Fokker-Planck synthetic acceleration (FPSA) - where the lower-order approximation for the error-correction stage is obtained using asymptotic analysis (Bender and Orszag, 1978) in the Fokker-Planck limit. We call this approach to acceleration asymptotics-based acceleration. \\

We organize the remainder of this paper as follows. In the next section, we introduce the FPSA and describe how we discretize the angular Laplacian term of the Fokker-Planck equation - weighted finite difference (Morel, 1985) and moment preserving discretization (Warsa and Prinja, 2012). In section 3, we present angularly-continuous and angularly-discrete Fourier analyses for FPSA and contrast them. Thenafter, in section 4, we present an efficiency study for FPSA screened Rutherford kernel (SRK) (Pomraning, 1992), the exponential kernel (EK) (Prinja et al., 1992) and Henyey-Greenstein kernel (HGK) (Henyey and Greenstein, 1941). We conclude this paper with a summary in section 5. \\

\section{Fokker-Planck Synthetic Acceleration}

The ideas presented here can be extended to problems with time and energy dependence,
in multi-dimensions, and in curvilinear coordinates, but our presentation takes place
in the context of steady-state, monoenergetic, one-dimensional Cartesian coordinates.
We use standard notation and assume cgs units (Lewis and Miller, 1984), such that
for a domain $z \in [z_0, z_1]$ the transport equation is
\begin{subequations}
\LEQ{trans-equation}
\begin{equation}
\LEQ{trans}
   \mu \dfrac{\partial}{\partial z} \psi(z,\mu) + \sigma_{t}(z) \psi(z, \mu)
= \int_{-1}^{1} d\mu^{\prime} \: \sigma_s(z,\mu_0) \psi(z, \mu^{\prime}) + q(z,\mu),
\end{equation}
with boundary conditions
\begin{equation}
\LEQ{trans-equation-bc}
\psi(z_0,\mu) = \psi_{L}(\mu) \quad \text{for $\mu>0$} \quad \text{and} \quad
\psi(z_1,\mu) = \psi_{R}(\mu) \quad \text{for $\mu<0$}.
\end{equation}
\end{subequations}
The cross section $\sigma_s(\mu_0)$ depends on
the cosine of the laboratory frame scattering angle,
$\mu_0 = \onp \! \cdot \! \on$, for a particle traveling with incident direction $\onp$,
and exiting after a scattering event in direction $\on$.
Typically, this dependence is represneted with an expansion in Legendre polynomials,
whose expansion coefficients are
\begin{equation}
\sigma_{s,l}(z) = \int_{-1}^{1} d\mu_0 \: \sigma_s(z, \mu_0) P_l(\mu_0).
\end{equation}
Assuming the expansion is truncrated at order $L$, and using the addition theorem
for the normalized spherical harmonics, Eq. (\REQ{trans}) becomes
\begin{equation}
\label{eq:treq}
   \mu \dfrac{\partial}{\partial z} \psi(z,\mu) + \sigma_{t}(z) \psi(z, \mu)
= \sum_{l=0}^{L} \dfrac{2l+1}{2} \sigma_{s,l}(z) P_{l}(\mu) \phi_{l}(z) + q(z,\mu),
\end{equation}
where the scalar flux moments are
\begin{equation}
\label{eq:phi-mom}
\phi_{l}(z) = \int_{-1}^{1} d\mu \: P_{l}(\mu) \psi(z, \mu).
\end{equation}
Under certain restrictions on the scattering cross section and its expansion,
the Fokker-Planck equation is an asymptotic limit of the Boltzmann transport
equation when scattering is highly forward-peaked (Pomraning, 1992).
The slab geometry Fokker-Planck equation is
\begin{subequations}
\LEQ{fp-eq}
\begin{equation}
\LEQ{fp-equation}
  \mu \dfrac{\partial \psi}{\partial z} + \sigma_a(z) \psi(z,\mu)
- \dfrac{\sigma_{tr}(z)}{2} \lfp \psi(z,\mu) = q(z,\mu),
\end{equation}
where
\begin{equation}
\LEQ{fp-operator}
\lfp = \dfrac{\partial}{\partial \mu} (1-\mu^2) \dfrac{\partial}{\partial \mu}
\end{equation}
\end{subequations}
and the momentum transfer (or transport) cross section is $\sigma_{tr} = \sigma_{s,0} - \sigma_{s,1}$.
%---------------------------------------------
% Standard Solution and Synthetic Acceleration
%---------------------------------------------
\subsection{Standard Solution and FPSA}

We demonstrate the ideas behind source iteration and synthetic acceleration (Adams and Larsen, 2001) before describing FPSA. Source iteration is one of the simplest methods used to solve the Boltzmann equations. We begin by rewriting \EQ{treq}:
\begin{subequations}
\LEQ{trop}
\begin{equation}
\LEQ{trop-equation}
L\psi(z,\mu) = S\psi(z,\mu) + q(z,\mu),
\end{equation}
where
\begin{equation}
\LEQ{trop-operator}
L = \mu \frac{\partial}{\partial z} + \sigma_t(z) \quad \text{and} \quad S = \sum_{l=0}^{L} \dfrac{2l+1}{2} \sigma_{s,l}(z) P_{l}(\mu)\int_{-1}^{1} d\mu \: P_{l}(\mu).
\end{equation}
\end{subequations}
Source iteration for the Boltzmann equation is then written as:
\begin{equation}
\label{eq:trsi}
L\psi^{m+1}(z,\mu) = S\psi^m(z,\mu) + q(z,\mu),
\end{equation}
where $m$ is the iteration index. Fourier analysis of source iteration (Valougeorgis et al., 1988) returns $\frac{\sigma_{s,l}}{\sigma_t}$ as the eigenvalues of the iteration matrix. Depending on the scattering kernel, the spectral radius can approach unity. This could make solution extremely expensive. In order to optimize the performance of such schemes, the solution must be accelerated. The eigenvalues also suggest that for forward-peaked kernels, where $\frac{\sigma_{s,l}}{\sigma_t}$ decay slowly, accelerating higher order moments with nonzero magnitudes becomes essential. In order to understand synthetic acceleration (Kopp, 1963) consider \EQ{trsi}. We break the solution procedure as follows: \\
\begin{subequations}
\LEQ{synac}
\textbf{Predict}
\begin{equation}
\LEQ{predict}
L\psi^{m+\frac{1}{2}}(z,\mu) = S\psi^m(z,\mu) + q(z,\mu),
\end{equation}
\textbf{Correct}
\begin{equation}
\LEQ{correct}
\psi^{m+1}(z,\mu) = \psi^{m+\frac{1}{2}}(z,\mu) + F^{-1}S(\psi^{m+\frac{1}{2}}(z, \mu) - \psi^{m}(z, \mu))
\end{equation}
\textbf{Iterate if}
\begin{equation}
\LEQ{iterate}
||\phi_l^{m+1}(z) - \phi_l^{m}(z)||_\infty > tolerence
\end{equation}
\end{subequations}
Different choices of $F$ operator return different synthetic acceleration schemes. For example choosing $F$ as the diffusion operator returns DSA. For this paper, we choose $F$ as the Fokker-Planck operator:
\begin{equation}
\label{eq:fpop}
F = \mu \frac{\partial}{\partial z} + \sigma_a(z) - \frac{\sigma_{tr}(z)}{2}\lfp
\end{equation}
%---------------------------------------------
% Discretization
%---------------------------------------------
\subsection{Discretization}

In order to discretize the Boltzmann and FP equations, we use linear discontinuous finite element discretization (LD) in space (Warsa, 2014) and discrete ordinates (\Sn) in angle (Lewis and Miller, 1984). Moreover in order to discretize \lfp, we use weighted finite difference (WFD) (Morel, 1985) and moment preserving discretization (MPD) (Warsa and Prinja, 2012). For a more detailed presentation on spatial and angular discretization, we refer the readers to (Patel, 2016). For convenience, we briefly review angular discretization of FP equation based on (Morel, 1985) and (Warsa and Prinja, 2012). We use \Sn quadrature to discretize the FP equation \EQ{fp-eq} in angle by collocating the angular flux at the directions $\mu_n$
\begin{equation}
\LEQ{fp-sn}
  \mu \dfrac{\partial \psi_n}{\partial z} + \sigma_a(z) \psi_n(z) - \nfp_n \psi(z) = q_n(z),
\end{equation}
for $n=1,\ldots,N$.
A way to discretize the term $\nfp_n \psi(z)$, which denotes the discrete form of the angular
Laplacian operator Eq. (\REQ{fp-operator}) evaluated at angle $n$, has to be defined
in terms of the \Sn quadrature points and weights. \\

The three-point, WFD scheme for the angular Laplacian has weights that identical to those used
for the \Sn discretization of the one-dimensional spherical coordinates (Morel, 1985).
The WFD scheme is given by the following expressions.
\begin{equation}
\LEQ{fp-wdd}
  \nfp_n \psi(z) = \gamma_{n+1/2} \dot{\psi}_{n+1/2}(z) - \gamma_{n-1/2} \dot{\psi}_{n-1/2}(z),
\end{equation}
where
\begin{equation}
 \dot{\psi}_{n+1/2}(z) = \dfrac{\psi_{n+1}(z) - \psi_{n}(z)}{\mu_{n+1} - \mu_{n}},
\end{equation}
\begin{equation}
\LEQ{sph-weights}
\gamma_{n+1/2} = \gamma_{n-1/2} + \nu \mu_{n} w_{n}, \quad \text{with} \quad \gamma_{n-1/2} = 0.
\end{equation}
and where the \Sn quadrature normalization is $\nu$. \\
To develop the MPD method, we first recognize
that in one dimension the Legendre polynomials are eigenfunctions
of the Fokker-Planck operator
\begin{equation}
\LEQ{legendre}
\lfp P_l(\mu) = -l(l+1) P_l(\mu).
\end{equation}
For a twice-differentiable function $f(\mu)$, integrating twice by parts shows that the angular
Laplacian operator is self-adjoint with respect to the Legendre polynomials:
\begin{equation}
\LEQ{f-legendre}
\int_{-1}^{1} \left[\lfp P_l(\mu)\right] f(\mu) \, d\mu = \int_{-1}^{1} P_l(\mu) \left[\lfp f(\mu)\right] \, d\mu.
\end{equation}
Substituting \EQ{legendre} in \EQ{f-legendre}, the following integral relationship is readily obtained
\begin{equation}
\LEQ{e-legendre}
          \int_{-1}^{1} P_l(\mu) \left[\lfp f(\mu)\right] \, d\mu
= -l(l+1) \int_{-1}^{1} P_l(\mu) f(\mu) \, d\mu .
\end{equation}
We now evaluate this relationship with \Sn quadrature
for the angular flux $\psi_n(z)$ to get
\begin{equation}
\LEQ{fp-mat}
           \sum_{n=1}^{N} w_n P_l(\mu_n) \nfp_n \psi(z) \\
=  -l(l+1) \sum_{n=1}^{N} w_n P_l(\mu_n) \psi_n(z),
\end{equation}
for $\quad l=0,\ldots,N-1$.
This defines an $(N \times N)$ operator for the vector of $N$ angular fluxes at the spatial location
$z$, $\Psi(z)$, such that the result is the Fokker-Planck operator collocated at all $N$ quadrature points
simultaneously.
That is,
\begin{subequations}
\LEQ{fp-mpd}
\begin{equation}
\LEQ{fp-mpd-def}
\nfp \Psi(z) = \bA{F} \Psi(z)
\end{equation}
where $\nfp$ is the discrete approximation to $\lfp$,
and where
\begin{equation}
\bA{F} = \bA{V}^{-1} \bA{L} \bA{V},
\end{equation}
where the elements of $\bA{V}$ and $\bA{L}$ are
\begin{equation}
\begin{aligned}
\bA{V}_{i,j} &= P_{i-1}(\mu_j) w_j  \\
\bA{L}_{i,i} &= -i(i-1),
\end{aligned}
\end{equation}
for $i,j = 1,\ldots,N$.
\end{subequations}
The method is a similarity transformation that by definition preserves
the moments of the flux up to order $N-1$.
Using Eq. (\REQ{fp-mpd}), the MPD method for the \Sn approximation to the Fokker-Planck
equation is, in operator notation,
\begin{equation}
\LEQ{fp-mpd-oper}
\bA{H} \dfrac{\partial}{\partial z} \Psi(z) + \sigma_a(z) \Psi(z)
- \dfrac{\sigma_{tr}(z)}{2} \bA{F} \Psi(z) = Q(z),
\end{equation}
where the operator $\bA{M}$ is
\[
\bA{H} = \mathop{\text{diag}}_{n=1,N}(\mu_n),
\]
and $Q(z)$ is a vector of source terms $q_n(z)$ for $n=1,\ldots,N$. \\

Notice that we may also write the WFD operator in the manner of Eq. (\REQ{fp-mpd-def}),
that is,
\begin{subequations}
\LEQ{fp-wd}
\begin{equation}
\LEQ{fp-wd-def}
\nfp \Psi(z) = \bA{W} \Psi(z)
\end{equation}
where $\bA{W}$ is the $(N \times N)$ tridiagonal matrix
whose elements are
\begin{equation}
\label{w_wfd}
\bA{W}_{i,j} =
\begin{cases}
\begin{aligned}
&\dfrac{1}{w_n} \left(\dfrac{\gamma_{n-1/2}}{\mu_n - \mu_{n-1}}\right), \quad &j&=i-1, \, j > 1, \\
&\dfrac{1}{w_n} \left(\dfrac{\gamma_{n+1/2}}{\mu_{n+1} - \mu_n}\right), \quad &j&=i+1, \, j < N, \\
&-\dfrac{1}{w_n} \left(\dfrac{\gamma_{n-1/2}}{\mu_n - \mu_{n-1}} + \dfrac{\gamma_{n+1/2}}{\mu_{n+1} - \mu_n}\right), \quad &j&=i,
\end{aligned}
\end{cases}
\end{equation}
\end{subequations}
for $i,j=1,\ldots,N$.
The WFD scheme for the \Sn approximation to the Fokker-Planck equation can then be written
in operator notation as
\begin{equation}
\LEQ{wd-eq-oper}
\bA{H} \dfrac{\partial}{\partial z} \Psi(z) + \sigma_a(z) \Psi(z)
- \dfrac{\sigma_{tr}(z)}{2} \bA{W} \Psi(z) = Q(z),
\end{equation}
As observed in (Morel, 1985), we see that the WFD scheme results
in a diagonally-dominant M-matrix such that the transport operator is inverse-positive
(neglecting spatial discretization).
Even though the MPD operator does not have a similar simple structure that allows us to
show it so easily, we have observed numerically that it is in fact inverse-positive. \\

\section{Fourier Analysis for FPSA}

Now that we have some idea of what FPSA is and how we discretize equations, we analyze FPSA using angularly-continuous and angularly-discrete Fourier analysis in this section. The goal of fourier analysis is to theoretically deretmine how efficient our method is. We begin by determining the error eqution for FPSA. We assume constant material properties throughout this exercise. We also drop notation for $z$ and $\mu$ dependence henceforth for conveneince. 

Consider Eq. \REQ{correct}. Upon subtracting the exact solution $\psi$ from both sides and adding and subtracting $\psi$ from the scattering term, we have:
\begin{equation}
\label{fa2}
\psi^{m+1} - \psi = \psi^{m+\frac{1}{2}} - \psi + F^{-1}S(\psi^{m+\frac{1}{2}} -\psi + \psi - \psi^m).
\end{equation}
Upon introduction of the following error definitions in Eq. \eqref{fa2}:
\begin{equation}
\label{err_defs}
\psi - \psi^{m+ \frac{1}{2}} = \epsilon^{m+\frac{1}{2}}, \quad
\psi - \psi^{m+\frac{1}{2}} = \epsilon^{m+\frac{1}{2}}, \quad \mathrm{and} \quad
\psi - \psi^{m} = \epsilon^{m},
\end{equation}
simplification, and rearrangement, we get the following error equation:
\begin{equation}
\label{err_fa1}
\epsilon^{m+1} = \epsilon^{m + \frac{1}{2}} - F^{-1}S(\epsilon^m - \epsilon^{m+\frac{1}{2}}).
\end{equation}
Based on whether we analyze the equation via Legendre-moments of angular error or by discretizing angular error with \Sn quadrature, we get angularly-continuous or angularly-discrete analysis. 

\subsection{Angularly-Continuous Fourier Analysis}

Angularly-continuous or \Pl-based Fourier analysis is inspired by (Valougeorgis et al., 1988). We begin by defining error moments:
\begin{equation}
\label{err_moment}
\epsilon_{l}^{m} = \int \limits_{-1}^{1} d\mu P_l(\mu)\epsilon^m = \sum_{n=1}^N w_n P_l(\mu_n)\epsilon^m.
\end{equation}
In terms of moments, the error equation is:
\begin{equation}
\label{moment_err}
\epsilon_l^{m+1} = \epsilon_l^{m + \frac{1}{2}} - \int_{-1}^1 d\mu P_l(\mu) F^{-1}S(\epsilon^m - \epsilon^{m+\frac{1}{2}}).
\end{equation}
We follow the following general steps:
\begin{enumerate}

\item Obtain an expression for $\epsilon_l^{m+\frac{1}{2}}$ by analying the predictor step.

\item Obtain an expression for $\int_{-1}^1 d\mu P_l(\mu) F^{-1}S(\epsilon^m - \epsilon^{m+\frac{1}{2}})$ by analyzing the corrector step.

\item Combine results from previous steps to obtain the iteration matrix $I_M$ such that the error is written according to $[\epsilon_l^{m+1}] = I_M [\epsilon_l^m]$.
\end{enumerate}

The spectral radius of $I_M$ determines the convergence rate of the iterative method (Hageman and Young, 1984). This is because of the following relation:
\begin{equation}
\label{epss}
[\epsilon_l^{m+1}] = I_M [\epsilon_l^m] = I_M^m [\epsilon^0].
\end{equation}  
\textbf{Step 1:} In order to proceed, we note that $\epsilon_l^{m+\frac{1}{2}}$ comes from the error moment equation of the predictor step Eq. (\REQ{predict}). We obtan that equation by subtracting the exact transport equation from Eq. (\REQ{predict}), and defining error according to Eq. \eqref{err_defs}: 
\begin{equation}
\label{errsi}
\mu \frac{\partial \epsilon^{m+\frac{1}{2}}}{\partial z} + \sigma_t \epsilon^{m+\frac{1}{2}} = \sum_{l=0}^{L} \frac{2l+1}{2}P_l(\mu)\sigma_{s,l} \int \limits_{-1}^{1} d\mu' P_l(\mu')\epsilon^{m}.
\end{equation}
Now, we separate the error components into their angle and space dependent components by writing $\epsilon^{m+\frac{1}{2}}$ and $\epsilon^m$ as Fourier integral (Adams and Larsen, 2001):
\begin{equation}
\epsilon^{m+\frac{1}{2}} = \int_{-\infty}^{\infty} d\lambda  \hat{\epsilon}_{\lambda}^{m+1}(\mu) e^{i \lambda \sigma_t z},
\end{equation}
where, $\lambda$ is the wave number. Substituting this form of error into the error equation Eq. \eqref{errsi} returns:
\begin{equation}
\label{errsi0}
\int_{-\infty}^{\infty} d\lambda \left(\mu \frac{\partial \hat{\epsilon}_{\lambda}^{m+1}(\mu) e^{i \lambda \sigma_t z}}{\partial z} + \sigma_t \hat{\epsilon}_{\lambda}^{m+1}(\mu) e^{i \lambda \sigma_t z} = \sum_{l=0}^{L} \frac{2l+1}{2}P_l(\mu)\sigma_{s,l} \int \limits_{-1}^{1} d\mu' P_l(\mu') \hat{\epsilon}_{\lambda}^{m}(\mu) e^{i \lambda \sigma_t z} \right).
\end{equation}
Simplifying the above equation and noting that Fourier modes, $e^{i \lambda \sigma_t z}$, are linearly independent for all $\lambda$, we obtain (Adams and Larsen, 2001):
\begin{equation}
\label{fa_si1}
(1 + i \lambda \mu) \sigma_t \hat{\epsilon}_{\lambda}^{m+1}(\mu) = \sum_{l=0}^{L} \frac{2l+1}{2}P_l(\mu)\sigma_{s,l} \int \limits_{-1}^{1} d\mu' P_l(\mu') \hat{\epsilon}_{\lambda}^{m}(\mu)
\end{equation}
Dropping $\lambda$ and $\mu$ in notation of $\hat{\epsilon}$ in Eq. \eqref{fa_si1} for convenience, and using definitions of error-moments, we get:
\begin{equation}
\label{fa_si2}
(1 + i \lambda \mu \sigma_t) \sigma_t \hat{\epsilon}^{m+\frac{1}{2}} = \sum_{l=0}^{L} \frac{2l+1}{2}P_l(\mu)\sigma_{s,l} \hat{\epsilon}_l^{m}.
\end{equation}
Then rearranging the equation and taking $n^{th}$ Legendre moment of Eq. \eqref{fa_si2}, we obtain the following:
\begin{equation}
\label{fa_si3}
\int_{-1}^1 d\mu P_n(\mu) \hat{\epsilon}_{\lambda}^{m+\frac{1}{2}} =  \int_{-1}^1 d\mu P_n(\mu) \sum_{l=0}^L \frac{\sigma_{s,l}}{\sigma_t} \frac{2l+1}{2} \frac{P_l(\mu)}{1 + i \lambda \mu \sigma_t}\hat{\epsilon}_l^{m}.
\end{equation}
Further rearrangement and use of definition of error moments returns:
\begin{equation}
\label{fa_si4}
\hat{\epsilon}_l^{m+\frac{1}{2}} = \sum_{l=0}^L \frac{\sigma_{s,l}}{\sigma_t} \frac{2l+1}{2} \int_{-1}^1 d\mu \frac{P_n(\mu)P_l(\mu)}{1 + i \lambda \mu \sigma_t}\hat{\epsilon}_l^{m}.
\end{equation}
We note that Eq. \eqref{fa_si4} represents the following matrix equation:
\begin{equation}
\label{fa_si5}
[\hat{\epsilon}_l^{m+\frac{1}{2}}] = A [\hat{\epsilon}_l^{m}],
\end{equation}
where, $[\hat{\epsilon}_l^{m}]$ is a vector of error-moments at iteration m, and
\begin{equation}
\label{fa_si6}
A = \sum_{l=0}^L \frac{\sigma_{s,l}}{\sigma_t} \frac{2l+1}{2} \int_{-1}^1 d\mu \frac{P_n(\mu)P_l(\mu)}{1 + i \lambda \mu \sigma_t},
\end{equation}
is an iteration matrix. We multiply Eq. \eqref{fa_si5} by $e^{i \lambda \sigma_t z}$ and use Eq. \eqref{err_moment} to get:
\begin{equation}
\label{fsa4}
[\epsilon_l^{m+\frac{1}{2}}] = A[\epsilon_l^{m}].
\end{equation}
Now that we have an equation for $\epsilon_l^{m+\frac{1}{2}}$, we move on to the next step. \\

\textbf{Step 2:} We begin from Eq. \eqref{err_fa1}. We note that the correction $\upsilon^{m+1} = F^{-1}S(\epsilon^m - \epsilon^{m+\frac{1}{2}})$ comes from the solution of the following equation:
\begin{equation}
\label{fpl_err}
\mu \frac{\partial \upsilon^{m+1}}{\partial z} + \sigma_a \upsilon^{m+1} - \frac{\sigma_{tr}}{2} \frac{\partial}{\partial \mu} (1-\mu^2) \frac{\partial \upsilon^{m+1}}{\partial \mu} = \sum_{l=0}^L \frac{2l+1}{2}P_l(\mu)\sigma_{s,l}(\epsilon_l^m - \epsilon_l^{m+\frac{1}{2}})
\end{equation}
Introducing the Fourier mode ansatz:
\begin{equation}
\label{fm1}
\upsilon^{m+1} = \hat{\upsilon}_{\lambda}^{m+1}(\mu)e^{i\lambda z \sigma_t}.
\end{equation}
Upon introduction of Eq. \eqref{err_moment}, and Eq. \eqref{fm1} in Eq. \eqref{fpl_err}, we get:
\begin{equation}
\label{fpl_err_1}
\begin{split}
\mu \frac{\partial \hat{\upsilon}_{\lambda}^{m+1}(\mu)e^{i\lambda z \sigma_t}}{\partial z} + \sigma_a \hat{\upsilon}_{\lambda}^{m+1}(\mu)e^{i\lambda z \sigma_t} - \frac{\sigma_{tr}}{2} \frac{\partial}{\partial \mu} (1-\mu^2) \frac{\partial \hat{\upsilon}_{\lambda}^{m+1}(\mu)e^{i\lambda z \sigma_t}}{\partial \mu} \\
= \sum_{l=0}^L \frac{2l+1}{2}P_l(\mu)\sigma_{s,l}(e^{i \lambda \sigma_t z}\hat{\epsilon}_l^{m} - e^{i \lambda \sigma_t z}\hat{\epsilon}_l^{m+\frac{1}{2}}).
\end{split}
\end{equation}
Simplifying Eq. \eqref{fpl_err_1}, taking its Legendre moment, and using the orthogonality property of Legendre polynomials returns:
\begin{equation}
\label{fpl_err1}
\begin{split}
i\lambda\sigma_t \int \limits_{-1}^1 d\mu P_l(\mu)\mu \hat{\upsilon}_{\lambda}^{m+1}(\mu) + \sigma_a \int \limits_{-1}^1 d\mu P_l(\mu) \hat{\upsilon}_{\lambda}^{m+1}(\mu) - \frac{\sigma_{tr}}{2} \int \limits_{-1}^1 d\mu \frac{\partial}{\partial \mu} (1-\mu^2) \frac{\partial \hat{\upsilon}_{\lambda}^{m+1}(\mu)}{\partial \mu} \\ = \sigma_{s,l} (\hat{\epsilon}_l^{m} - \hat{\epsilon}_l^{m+ \frac{1}{2}}).
\end{split}
\end{equation}
Now, using the recurrence relation for Legendre polynomials on the first term of Eq. \eqref{fpl_err1}, expanding $\hat{\upsilon}_{\lambda}^{m+1}(\mu)$ in the third term of Eq. \eqref{fpl_err1} using Legendre expansion, we get:
\begin{equation}
\label{fpl_err2}
\begin{split}
\frac{l}{2l+1} i\lambda \sigma_t \hat{\upsilon}_{l-1}^{m+1} + \frac{l+1}{2l+1} i\lambda \sigma_t \hat{\upsilon}_{l+1}^{m+1} + \sigma_a \hat{\upsilon}_{l}^{m+1} - \frac{\sigma_{tr}}{2}  \int \limits_{-1}^1 d\mu \frac{\partial}{\partial \mu} (1-\mu^2) \frac{\partial}{\partial \mu} \sum_{n=0}^{\infty} \frac{2l+1}{2} P_n(\mu) \hat{\upsilon}_{n}^{m+1} \\ = \sigma_{s,l} (\hat{\epsilon}_l^{m} - \hat{\epsilon}_l^{m+ \frac{1}{2}}).
\end{split}
\end{equation}
Simple rearrangement of the third term in Eq. \eqref{fpl_err2}, followed by use of Legendre's equation, and orthogonality property of Legendre polynomials returns:
\begin{equation}
\label{fpl_err3}
\frac{l}{2l+1} i\lambda \sigma_t \hat{\upsilon}_{l-1}^{m+1} + \frac{l+1}{2l+1} i\lambda \sigma_t \hat{\upsilon}_{l+1}^{m+1} + \sigma_a \hat{\upsilon}_{l}^{m+1} + \frac{\sigma_{tr}}{2} l(l+1) \hat{\upsilon}_{n}^{m+1}  = \sigma_{s,l} (\hat{\epsilon}_l^{m} - \hat{\epsilon}_l^{m+ \frac{1}{2}}),
\end{equation}
where,
\begin{equation}
\label{up_mom}
\hat{\upsilon}_l^{m} = \int \limits_{-1}^{1} d\mu P_l(\mu) \hat{\upsilon}_{\lambda}^{m}(\mu).
\end{equation}
Eq. \eqref{fpl_err3} can be written in matrix form as:
\begin{equation}
\label{fpl_err4}
[\hat{\upsilon}_{l}^{m+1}] = B^{-1}XE[\hat{\epsilon}_l^m]
\end{equation}
where,
\begin{subequations}
\label{ops}
\begin{equation}
X = diag(\sigma_{s,l}),
\end{equation}
\begin{equation}
E = I - A
\end{equation}
\begin{equation}
B_{l,l} = \sigma_a + \frac{\sigma_{tr}}{2} l(l+1)
\end{equation}
\begin{equation}
B_{l,l+1} = \frac{l+1}{2l+1} i \lambda \sigma_t
\end{equation}
\begin{equation}
B_{l-1,l} = \frac{l}{2l+1} i \lambda \sigma_t
\end{equation}
\end{subequations}
Multiplying Eq. \eqref{fpl_err4} with $e^{i\lambda z \sigma_t}$ and using Eq. \eqref{fm1} returns:
\begin{equation}
\label{fpl_err5}
[\upsilon_{l}^{m+1}] = B^{-1}XE[\epsilon_l^m] = \int_{-1}^1 d\mu P_l(\mu) F^{-1}S(\epsilon^m - \epsilon^{m+\frac{1}{2}}).
\end{equation}

\textbf{Step 3:} Combining Eqs. \eqref{moment_err}, \eqref{fsa4}, and \eqref{fpl_err5}, returns:
\begin{equation}
\label{fpl_err6}
[\epsilon_l^{m+1}] = (A-B^{-1}XE)[\epsilon_l^{m}].
\end{equation}
Comparing Eq. \eqref{fpl_err6} with Eq. \eqref{epss} returns the iteration matrix $I_M = (A-B^{-1}XE)$. The spectral radius of $I_M$ is the spectral radius of FPSA.

\subsection{FPSA as a Special Case of \Pl Acceleration}

Upon carrying out similar analysis for \Pl acceleration (Valougeorgis et al., 1988), we find that the iteration matrix for \Pl acceleration has a similar form except, in this case, the definition of $B_{l,l}$ is slightly different:
\begin{equation}
\label{sa_diff}
B_{l,l}^{P_L} = \sigma_t - \sigma_{s,l} = \sigma_a + \sigma_{s,0} - \sigma_{s,l}.
\end{equation}
That for FPSA is rewritten as: 
\begin{equation}
B_{l,l}^{FPSA} = \sigma_a + \frac{\sigma_{tr}}{2} l(l+1) = \sigma_a + \frac{\sigma_{s,0} - \sigma_{s,1}}{2} l(l+1).
\end{equation}
When we equate the two equations, we see that FPSA is a special case of \Pl acceleration when:
\begin{equation}
\label{fsa_pl_eq}
\sigma_{s,l} = \sigma_{s,0} - \frac{\sigma_{s,0} - \sigma_{s,1}}{2}l(l+1).
\end{equation}
Another way of obtaining this equivalence relation is by noting that Legendre polynomials are eigenfunctions of both Boltzmann scattering operator and the Fokker-Planck operator as done by Morel in (Morel, 1981):
\begin{subequations}
\label{subs1}
\begin{equation}
\Gamma_B P_l(\mu) = (\sigma_{s,l} - \sigma_{s,0})P_l(\mu),
\end{equation}
\begin{equation}
\Gamma_{FP} P_l(\mu) = -\frac{(\sigma_{s,0} - \sigma_{s, 1})}{2}l(l+1)P_l(\mu),
\end{equation}
\end{subequations}
and equating the eigenvalues of Fokker-Planck and the Boltzmann scattering operators:
\begin{equation}
\label{fp_tr_eq}
\sigma_{s,l} - \sigma_{s,0} = -\frac{(\sigma_{s,0} - \sigma_{s, 1})}{2}l(l+1).
\end{equation}
Simple rearrangement of Eq. \eqref{fp_tr_eq} returns Eq. \eqref{fsa_pl_eq}. We will call these cross-section moments $P_L$-equivalent cross-section moments in this paper. \\

According to the $S_N-P_L$ equivalence relation in slab geometry (Lewis and Miller, 1984), when $N = L+1$, $S_N$ and \Pl equations are equivalent. Taking this and Eq. \eqref{fsa_pl_eq} into account, we note that FPSA will converge in one iteration when it is analytically equivalent to \Pl acceleration. In other words, when the scattering cross-section moments are according to Eq. \eqref{fsa_pl_eq}, FPSA will converge in one iteration. Moreover, it would be a valid to think that the convergence will be rapid in case the cross-section moments are close to those obtained from  Eq. \eqref{fsa_pl_eq}. However, in the case when we truncate scattering expansion arbitrarily and $N$ is no longer equal to $L+1$, the FPSA-\Pl acceleration equivalence will no longer hold. This is due to the inconsistent introduction of zero values for scattering crosssection moments with $N \geq l > L$ (Patel, 2016). 

\subsection{Angularly-Discrete Fourier Analysis}

Angularly-discrete analysis is carried out by using \Sn quadrature to approximate angular error. We need angularly-discrete Fourier analysis to analyze FPSA because different discretizations of $\lfp$ preserve different number of moments. While WFD only preserves zeroth and first Legendre moments of the angular flux (Morel, 1985), MPD preserves upto $N$ Legendre moments (Warsa and Prinja, 2012). The angluarly-continuous Fourier analysis (\Pl-based analysis) is moment-based and therefore requires the numerical implementation to preserve all relevant moments in order to get a consistent spectral radius measurement. Moreover, in case of continuous transport, the transport equation only limits to the Fokker-Planck equation when a "sufficient" number of Legendre moments are used to represent the angular flux (Patel, 2016). This sufficient number of moments is scattering cross-section dependent. This, however, may not necessarily be true in the discrete $(S_N)$ case. This creates a discrepancy between the angularly-continuous and angularly-discrete Fourier analyses when sufficient number of moments are not used to represent angular flux. Therefore, in order to verify convergence rates of the numerical implementation, irrespective of how many moments are used to represent angular flux, we introduce angularly-discrete analysis. First, we consider Fourier analysis for FPSA with WFD for $\lfp$.

\subsubsection{Analysis with WFD}

We follow the following analogous steps to do angularly discrete Fourier analysis:

\begin{enumerate}
\item Obtain an expression for $\epsilon^{m+\frac{1}{2}}$.
\item Obtain an expression for $F^{-1}S(\epsilon^{m} - \epsilon^{m+\frac{1}{2}})$.
\item Obtain the overall matrix equation that is used to estimate the the spectral radius.
\end{enumerate}

\textbf{Step 1:} Since we have already detailed angularly-continuous analysis, we skip furnishing the introduction of Fourier mode assumption and simplification steps here. We also ignore notation of $\mu$ and $z$ dependence of relevant quantities for convenience. Taking the $n^{th}$ Legendre moment of Eq. \eqref{fa_si2}, and using orthogonality property of Legendre polynomial returns:
\begin{equation}
\label{fpsa3}
\int \limits_{-1}^1 d\mu P_n(\mu) \left[(i\lambda\sigma_t\mu + \sigma_t)\hat{\epsilon}_l^{m+\frac{1}{2}} \right] = \sigma_{s,l}\int \limits_{-1}^1 P_l(\mu)\hat{\epsilon}_l^m .
\end{equation}
Now we write each integral as a discrete weighted-sum using \Sn quadrature:
\begin{equation}
\label{fpsa4}
\sum_{n=1}^N P_l(\mu_n)w_n \left[(i\lambda\sigma_t\mu_n + \sigma_t)\hat{\epsilon}_l^{m+\frac{1}{2}} \right] = \sigma_{s,l} \sum_{n=1}^N P_l(\mu_n)w_n\hat{\epsilon}_l^m.
\end{equation}
Finally, we get the following matrix equation from Eq. \eqref{fpsa4}:
\begin{equation}
\label{fpsa5}
[\hat{\epsilon}^{m+\frac{1}{2}}] = \hat{A}[\hat{\epsilon}^m].
\end{equation}
where,
\begin{equation}
\label{fpsa6}
\hat{A} = Y^{-1}Z,
\end{equation}
and,
\begin{equation}
\label{fpsa7}
Y_{ln} = P_l(\mu_n)w_n(i\lambda\sigma_t\mu_n + \sigma_t) \quad \mathrm{and} \quad Z_{ln} = \sigma_{s,l}P_l(\mu_n)w_n.
\end{equation}
Here, $\hat{A}$ is the iteration matrix in angularly-discrete from. This returns $\frac{\sigma_{s,0}}{\sigma_t}$ as the spectral radius which is consistent with the angularly-continuous analysis (Patel, 2016). We multiply Eq. \eqref{fpsa5} by the relevant exponential from Fourier mode ansatz to get: 
\begin{equation}
\label{fpsa5}
[\epsilon^{m+\frac{1}{2}}] = \hat{A}[\epsilon^m].
\end{equation}
\textbf{Step 2:} Upon introduction of Fourier mode assumption for $\upsilon$ in Eq. \eqref{fpl_err}, taking Legendre moment of equation, using definition of error moments, carrying out the relevant spatial differentiation and simplifications, we get:
\begin{equation}
\label{fper2}
\int \limits_{-1}^1 d \mu P_l(\mu) \left[i\lambda\sigma_t\mu + \sigma_a - \frac{\sigma_{tr}}{2} \frac{\partial}{\partial \mu} (1-\mu^2) \frac{\partial}{\partial \mu} \right]\hat{\upsilon}^{m+1} = \sigma_{s,l} \int \limits_{-1}^1 d \mu P_l(\mu)(\hat{\epsilon}^m - \hat{\epsilon}^{m+\frac{1}{2}}).
\end{equation}
Now we write each integral in the form of a weighted sum and the angular differential using the weighted difference formulation (Morel, 1989) to obtain:
\begin{multline}
\label{fper3}
i\lambda\sigma_t \sum_{n=1}^N P_l(\mu_n)w_n\hat{\upsilon}_n^{m+1} + \sigma_a \sum_{n=1}^N P_l(\mu_n)w_n\hat{\upsilon}_n^{m+1} \\ - \frac{\sigma_{tr}}{2} \sum_{n=1}^N P_l(\mu_n) w_n \left(a_n \hat{\upsilon}_{n+1}^{m+1} - b_n \hat{\upsilon}_n^{m+1} + c_n \hat{\upsilon}_{n-1}^{m+1} \right) = \sigma_{s,l} \sum_{n=1}^N P_l(\mu_n) w_n (\hat{\epsilon}^m - \hat{\epsilon}^{m+\frac{1}{2}}),
\end{multline}
where, $a_n, b_n,$ and $c_n$ are according to Eq. \eqref{w_wfd}. From Eq. \eqref{fper3}, we get the following matrix equation:
\begin{equation}
\label{fper}
[\hat{\upsilon}^{m+1}] = \hat{B}^{-1}\hat{C}\hat{D}[\hat{\epsilon}^m],
\end{equation}
where,
\begin{equation}
\label{matdef1}
\hat{D} = I-\hat{A}, \quad \hat{C}_{l,n} = \sigma_{s,l} P_l(\mu_n) w_n, \quad \mathrm{and} \quad \hat{B} = \hat{B1} + \hat{B2} + \hat{B3},
\end{equation}
with,
\begin{equation}
\label{matdef2}
\hat{B1}_{l,n} =  P_l(\mu_n) w_n \left(i\lambda\sigma_t\mu_n + \sigma_a + b_n \right),
\end{equation}
\begin{equation}
\label{matdef3}
\hat{B2}_{l,{n+1}} =  P_l(\mu_n) w_n a_n,
\end{equation}
and,
\begin{equation}
\label{matdef3_1}
\hat{B2}_{l,{n-1}} =  P_l(\mu_n) w_n c_n.
\end{equation}
We obtain the following expression for $F^{-1}S(\epsilon^{m} - \epsilon^{m+\frac{1}{2}})$:,
\begin{equation}
\label{fperr}
\upsilon^{m+1} = \hat{\upsilon}^{m+1}e^{i\lambda\sigma_tz} = F^{-1}S(\epsilon^m - \epsilon^{m+\frac{1}{2}}).
\end{equation}

\textbf{Step 3:} Combining Eqs. \eqref{matdef3_1}, \eqref{fpsa5}, and \eqref{err_fa1} returns:
\begin{equation}
[\epsilon^{m+1}] = (\hat{A}-\hat{B}^{-1}\hat{C}\hat{D})[\epsilon^m],
\end{equation}
where, $I_M = \hat{A}-\hat{B}^{-1}\hat{C}\hat{D}$ is the iteration matrix and its spectral radius determines the convergence rate of FPSA with WFD. Now, we consider angularly-discrete analysis for FPSA with MPD.

\subsubsection{Analysis with MPD}

Angularly discrete Fourier analysis for FPSA with MPD is done in the same way as for FPSA with WFD. The only difference will be how the Fokker-Planck operator is represented in step 2. Introducing angularly discrete formulation for integrals and MPD formulation (Warsa and Prinja, 2012) for the angular Laplacian in Eq. \eqref{fper2} returns:
\begin{multline}
\label{fper3mpd}
i\lambda\sigma_t \sum_{n=1}^N P_l(\mu_n)w_n\hat{\upsilon}_n^{m+1} + \sigma_a \sum_{n=1}^N P_l(\mu_n)w_n\hat{\upsilon}_n^{m+1} + \frac{\sigma_{tr}}{2} l(l+1) \sum_{n=1}^N P_l(\mu_n) w_n \hat{\upsilon}_n^{m+1}  \\ = \sigma_{s,l} \sum_{n=1}^N P_l(\mu_n) w_n (\hat{\epsilon}^m - \hat{\epsilon}^{m+\frac{1}{2}}).
\end{multline}
From Eq. \eqref{fper3}, we get the following matrix equation:
\begin{equation}
\label{fpermpd}
[\hat{\upsilon}^{m+1}] = \tilde{B}^{-1}\tilde{C}\tilde{D}[\hat{\epsilon}^m],
\end{equation}
where,
\begin{equation}
\label{matdef1mpd}
\tilde{D} = I-\hat{A},
\end{equation}
\begin{equation}
\label{matdef2mpd}
\tilde{C}_{l,n} = \sigma_{s,l} P_l(\mu_n) w_n,
\end{equation}
and,
\begin{equation}
\label{matdef3mpd}
\tilde{B}_{l,n} = i \lambda \sigma_t w_n P_l(\mu_n) \mu_n + \sigma_a  w_n P_l(\mu_n) + l(l+1) \frac{\sigma_{tr}}{2} w_n P_l(\mu_n).
\end{equation}
We have the following expression for $ L^{-1}S(\epsilon^m - \epsilon^{m+\frac{1}{2}})$:
\begin{equation}
\label{fperr_1}
\upsilon^{m+1} = \hat{\upsilon}^{m+1}e^{i\lambda\sigma_tz} = L^{-1}S(\epsilon^m - \epsilon^{m+\frac{1}{2}}).
\end{equation}

\textbf{Step 3:} Just like with previous analyses for FPSA, we get:
\begin{equation}
[\epsilon^{m+1}] = (\hat{A}-\tilde{B}^{-1}\tilde{C}\tilde{D})[\epsilon^m].
\end{equation}
Thus the iteration matrix for FPSA with MPD is $ A-B^{-1}CD $. 

\subsection{Comparison of Spectral Radii}

In order to get a glimpse into how FPSA performs, we consider one problem with screened Rutherford kernel (SRK) (Dixon, 2015), exponential kernel (EK) (Prinja et al., 1992) and Henyey-Greenstein kernel (HGK) (Pomraning, 1992) each. We plot spectral radii of stand-alone \Sn and FPSA for each kernel in the figures that follow. We choose $L = 15$, $N = 16$. We note that the spectral radius reduces significantly upon introduction of FPSA. The spectral radius reduction is completely problem dependent. The spectral radius can potentially change with N, L, and how close the scattering cross-section moments of the problem are to the $P_L$-equivalent moments. 
\begin{figure}[H]
\begin{center}
\includegraphics[scale=0.8]{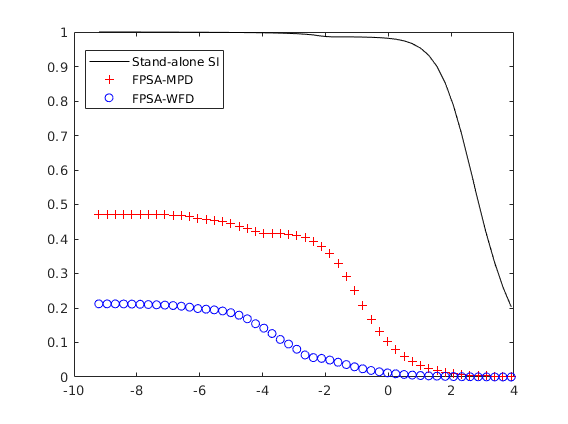}
\caption{Comparison of Source Iteration and FPSA - SRK - $\eta = 2.836 \times 10^{-5}$}
\label{fig:srk}
\end{center}
\end{figure}
\begin{figure}[H]
\begin{center}
\includegraphics[scale=0.8]{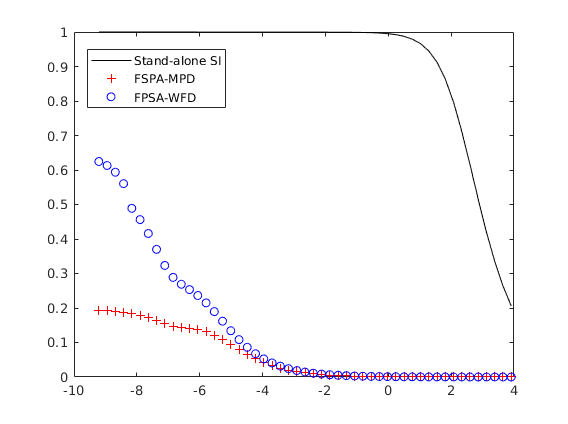}
\caption{Comparison of Source Iteration and FPSA - EK - $\Delta = 10^{-5}$}
\label{fig:ek}
\end{center}
\end{figure}
\begin{figure}[H]
\begin{center}
\includegraphics[scale=0.8]{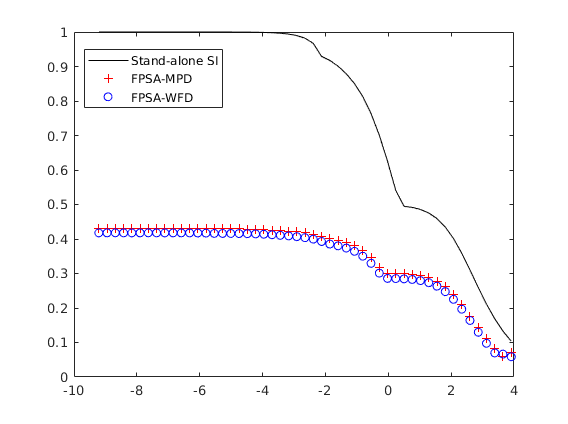}
\caption{Comparison of Source Iteration and FPSA - HGK - $g = 0.9$}
\label{fig:hgk}
\end{center}
\end{figure}
Next, we compare the numerically measured and theoretical (angularly-discrete Fourier analysis) spectral radii. We analyze convergence rates for three scattering kernels - SRK, EK, and HGK. We choose $L = 15$, $N = 16$.  We use a slab of length, 100 cm, discretize it using 100 elements. We use vacuum boundaries for numerical measurements of spectral radius. The theoretical and numerically measured spectral radii have been presented in Table \ref{table:sprcomp}. \\
\linespread{1}
\begin{table}[H]
\centering
\begin{tabular}{c c c c c}
\hline \hline
\textbf{Kernel/Parameter} & \textbf{$\mathrm{\rho_{FPSA}^{MPD}}$-FA} & \textbf{$\mathrm{\rho_{FPSA}^{MPD}}$-Measured} & \textbf{$\mathrm{\rho_{FPSA}^{WFD}}$-FA} & \textbf{$\mathrm{\rho_{FPSA}^{WFD}}$-Measured}  \\ 
\hline
SRK/$\eta = 2.83 \times 10^{-5}$ & 0.4706 & 0.4706 & 0.2121 & 0.2120 \\
%SRK/$\eta = 2.83 \times 10^{-6}$ & 0.3906 & 0.3898 & 0.3213 & 0.3215 \\
% EK/$\Delta = 10^{-4}$ & 0.2101 & 0.1975 & 0.6299 & 0.6301 \\
EK/$\Delta = 10^{-5}$ & 0.1932 & 0.1954 & 0.6246 & 0.6327 \\
HGK/$g = 0.9$ & 0.4304 & 0.4303 & 0.4177 & 0.4177 \\
%HGK/$g = 0.9999$ & 0.8709 & 0.8688 & 0.7439 & 0.7369 \\
\hline
\end{tabular}
\caption{Comparison of Numerical and Theoretical Spectral Radii}
\label{table:sprcomp}
\end{table}
\linespread{2}
We obtain similar theoretical and measured spectral radii values for different scattering kernels with varying parameters. This indicates a relatively accurate analysis of the method.
\section{Efficiency Study}

In this section we will assess how the reduction in spectral radius results in reduction in runtime of source iteration (SI) and GMRES solves. We run all problems using MATLAB and track runtime using its tic-toc functionality. We place tic and toc before and after the solver function calls respectively. In other words, we do not include the stiffness matrix setup time in our calculation. We will only account for the solver runtime. Specifically, choose problems with $L = 15$, and $N = 16$, $32$. We use beam and vacuum boundaries. We have a unit distributed source for problems with vacuum boundaries and a unit beam source with the beam boundary. We do this for SRK with $\eta = 2.83 \times 10^{-5}$, and for EK with $\Delta = 10^{-5}$. We solve the Fokker-Planck error equation (invert the preconditioner) using LU factorization via factorize object (Davis, 2009) in MATLAB, and GMRES. \\ 

First, we compare unpreconditioned SI and GMRES solves. In order to compare these solves, we choose $\eta = 2.83 \times 10^{-5}$, $L = 15$, $N = 16$, $H = 1$cm, $K = 100$, $tol = 10^{-10}$. We do this to contrast source iteration and GMRES solves. Table \ref{table:Unpreconditioned_iter_srk} and \ref{table:Unpreconditioned_time_srk} present this data. It is clear that GMRES is more suitable than source iteration for forward-peaked transport problems. \\

\linespread{1}
\begin{table}[H]
\centering
\begin{tabular}{c c c c}
\hline \hline
\textbf{BC/Source} & \textbf{Restart} & \textbf{GMRES Iteration Count} & \textbf{SI Iteration Count}  \\
\hline
Vacuum/Distributed & & &  $>$ 150000 \\
				   & 50 & 3305 & \\
				   & 100 & 2445 & \\
				   & 150 & 1875 & \\
				   & 200 & 1540 & \\
Beam/Zero		   &  &  & $>$ 150000 \\
 				   & 50 & 2602 & \\
				   & 100 & 2200 & \\
				   & 150 & 1895 & \\
				   & 200 & 1735 & \\
\hline
\end{tabular}
\caption{SRK - Number of Iterations}
\label{table:Unpreconditioned_iter_srk}
\end{table}
\linespread{2}

\linespread{1}
\begin{table}[H]
\centering
\begin{tabular}{c c c c}
\hline \hline
\textbf{BC/Source} & \textbf{Restart} & \textbf{GMRES Runtime} & \textbf{SI Runtime}  \\
\hline
Vacuum/Distributed &  &  & $>$ 3000 \\
				   & 50 & 64.97 &  \\
				   & 100 & 50.41 & \\
				   & 150 & 37.76 & \\
				   & 200 & 32.71 & \\
Beam/Zero		   &  &  & $>$ 3000 \\
				   & 50 & 50.68 & \\
				   & 100 & 43.99 & \\
				   & 150 & 41.29 & \\
				   & 200 & 36.11 & \\
\hline
\end{tabular}
\caption{SRK - Solver Runtime [s]}
\label{table:Unpreconditioned_time_srk}
\end{table}
\linespread{2}
Next, we compare solution rutimes and iteration counts. We will compare these for unpreconditioned GMRES, FPSA-preconditioned SI, and FPSA-preconditioned GMRES solves. We do not include unpreconditioned source iteration in this study because its ineffectiveness for relevant problems has already been demonstrated in Table \ref{table:Unpreconditioned_iter_srk} and \ref{table:Unpreconditioned_time_srk}. We will arbitrarily choose our restart parameter for this study to be 150.

\subsection{Screened Rutherford Kernel}

We compare efficiency of FPSA for problems involving SRK in this section. We choose a slab of unit length discretized using hundred elements. We choose $\eta = 2.83 \times 10^{-5}$ and $\sigma_a = 1$.  Scattering cross-section moments come from SRK and their numerical values can be found in (Patel, 2016). Finally, $L = 15$, and $N = 16$ and $32$. Number of iterations and overall runtime data has been presented in Table \ref{table:iter_comp_srk_v}, \ref{table:rt_comp_srk_v}, \ref{table:iter_comp_srk_b}, and \ref{table:rt_comp_srk_b}.  \\
\linespread{1}
\begin{table}[H]
\centering
\begin{tabular}{c c c c c c c}
\hline \hline
\textbf{FP-Solve} & \textbf{L/N} & \textbf{$\mathrm{GMRES}$} & \textbf{$\mathrm{FPSA_{GMRES}^{MPD}}$} & \textbf{$\mathrm{FPSA_{GMRES}^{WFD}}$} & \textbf{$\mathrm{FPSA_{SI}^{MPD}}$} & \textbf{$\mathrm{FPSA_{SI}^{WFD}}$} \\
\hline
GMRES & 15/16 & 1487 & 9 & 6 & 14 & 10 \\
					 & 15/32  & 1499 & 9 & 7 & 14 & 12 \\
Factorize & 15/16 & & 9 & 7 & 14 & 10 \\
					 &  15/32 & & 9 & 8 & 14 & 12 \\
\hline
\end{tabular}
\caption{SRK - Vacuum Boundaries/Unit Distributed Source - Number of Iterations}
\label{table:iter_comp_srk_v}
\end{table}
\linespread{2}
\linespread{1}
\begin{table}[H]
\centering
\begin{tabular}{c c c c c c c}
\hline \hline
\textbf{FP-Solve} & \textbf{L/N} & \textbf{$\mathrm{GMRES}$} & \textbf{$\mathrm{FPSA_{GMRES}^{MPD}}$} & \textbf{$\mathrm{FPSA_{GMRES}^{WFD}}$} & \textbf{$\mathrm{FPSA_{SI}^{MPD}}$} & \textbf{$\mathrm{FPSA_{SI}^{WFD}}$} \\
\hline
GMRES & 15/16 & 28.75 & 8.79 & 6.15 & 12.01 & 5.98 \\
					 & 15/32  & 28.12 & 36.76 & 18.36 & 54.13 & 19.77 \\
Factorize & 15/16 & & 1.62 & 2.45 & 0.3373 & 0.2501 \\
					  & 15/32  & & 2.75 & 4.73 & 0.4244 & 0.3392 \\
\hline
\end{tabular}
\caption{SRK - Vacuum Boundaries/Unit Distributed Source - Runtime [s]}
\label{table:rt_comp_srk_v}
\end{table}
\linespread{2}
%
% BEAM ===============================================================
%
\linespread{1}
\begin{table}[H]
\centering
\begin{tabular}{c c c c c c c}
\hline \hline
\textbf{FP-Solve} & \textbf{L/N} & \textbf{$\mathrm{GMRES}$} & \textbf{$\mathrm{FPSA_{GMRES}^{MPD}}$} & \textbf{$\mathrm{FPSA_{GMRES}^{WFD}}$} & \textbf{$\mathrm{FPSA_{SI}^{MPD}}$} & \textbf{$\mathrm{FPSA_{SI}^{WFD}}$} \\
\hline
GMRES & 15/16 & 1357 & 12 & 8 & 21 & 13 \\
					 & 15/32  & 1335 & 13 & 10 & 23 & 19 \\
Factorize & 15/16 & & 12 & 9 & 21 & 13 \\
					 &  15/32 & & 13 & 11 & 23 & 19 \\
\hline
\end{tabular}
\caption{SRK - Beam Source - Number of Iterations}
\label{table:iter_comp_srk_b}
\end{table}
\linespread{2}
\linespread{1}
\begin{table}[H]
\centering
\begin{tabular}{c c c c c c c}
\hline \hline
\textbf{FP-Solve} & \textbf{L/N} & \textbf{$\mathrm{GMRES}$} & \textbf{$\mathrm{FPSA_{GMRES}^{MPD}}$} & \textbf{$\mathrm{FPSA_{GMRES}^{WFD}}$} & \textbf{$\mathrm{FPSA_{SI}^{MPD}}$} & \textbf{$\mathrm{FPSA_{SI}^{WFD}}$} \\
\hline
GMRES & 15/16 & 26.54 & 11.39 & 6.975 & 9.69 & 5.403 \\
					 & 15/32  & 27 & 42.59 & 20.50 & 58.3 & 20.26 \\
Factorize & 15/16 & & 1.687 & 2.547 & 0.4475 & 0.3074 \\
					  & 15/32  & & 2.927 & 5.061 & 0.611 & 0.4828 \\
\hline
\end{tabular}
\caption{SRK - Beam Source - Runtime [s]}
\label{table:rt_comp_srk_b}
\end{table}
\linespread{2}
We observe a significant decrease (almost three orders of magnitude compared to unpreconditioned GMRES and five orders of magnitude compared to SI) in the number of transport-sweeps required for convergence due to preconditioning. We also observe a decrease in overall solver runtimes due to preconditioning when FP-solve is done using LU factorization (by upto two orders of magnitude compared to unpreconditioned GMRES). The FP-solve, however, can be extremely expensive and render this preconditioner ineffective with respect to problem's overall runtime if inefficient solvers are used. Here, the number of iterations required for one FP-solve using GMRES was of the same order as an unpreconditioned transport solve using GMRES.  It is imperative that we find an effective preconditioner for FP-solves. We are looking into this. The potential, however, of using FP as preconditioner for transport solves is amply evident from the data presented in this section. Next, we look at efficiency data for problems with the exponential kernel.

\subsection{Exponential Kernel}

We calculate scattering cross-section moments using EK for $\Delta = 10^{-5}$. The zeroth moment is calculated using SRK. The study is done using the same parameters as SRK except for scattering cross-section moments. Number of iterations and overall runtime data has been presented in Table \ref{table:iter_comp_ek_v}, \ref{table:rt_comp_ek_v}, \ref{table:iter_comp_ek_b}, and \ref{table:rt_comp_ek_b}. \\
\linespread{1}
\begin{table}[H]
\centering
\begin{tabular}{c c c c c c c}
\hline \hline
\textbf{FP-Solve} & \textbf{L/N} & \textbf{$\mathrm{GMRES}$} & \textbf{$\mathrm{FPSA_{GMRES}^{MPD}}$} & \textbf{$\mathrm{FPSA_{GMRES}^{WFD}}$} & \textbf{$\mathrm{FPSA_{SI}^{MPD}}$} & \textbf{$\mathrm{FPSA_{SI}^{WFD}}$} \\
\hline
GMRES & 15/16 & 2217 & 7 & 8 & 9 & 15 \\
	  & 15/32  & 2256 & 10 & 9 & 17 & 19 \\
Factorize & 15/16 & & 7 & 9 & 9 & 15 \\
		  &  15/32 & & 10 & 10 & 17 & 19 \\
\hline
\end{tabular}
\caption{EK - Vacuum Boundaries/Unit Distributed Source - Number of Iterations}
\label{table:iter_comp_ek_v}
\end{table}
\linespread{2}
\linespread{1}
\begin{table}[H]
\centering
\begin{tabular}{c c c c c c c}
\hline \hline
\textbf{FP-Solve} & \textbf{L/N} & \textbf{$\mathrm{GMRES}$} & \textbf{$\mathrm{FPSA_{GMRES}^{MPD}}$} & \textbf{$\mathrm{FPSA_{GMRES}^{WFD}}$} & \textbf{$\mathrm{FPSA_{SI}^{MPD}}$} & \textbf{$\mathrm{FPSA_{SI}^{WFD}}$} \\
\hline
GMRES & 15/16 & 51.73 & 10.32 & 13.34 & 11.91 & 18.49 \\
	  & 15/32  & 56.10 & 31.14 & 25.06 & 42.82 & 27.85 \\
Factorize & 15/16 & & 2.3754 & 4.5586 & 0.3762 & 0.454 \\
		  & 15/32  & & 4.155 & 8.463 & 0.6098 & 0.6424 \\
\hline
\end{tabular}
\caption{EK - Vacuum Boundaries/Unit Distributed Source - Runtime [s]}
\label{table:rt_comp_ek_v}
\end{table}
\linespread{2}
%
% BEAM ===============================================================
%
\linespread{1}
\begin{table}[H]
\centering
\begin{tabular}{c c c c c c c}
\hline \hline
\textbf{FP-Solve} & \textbf{L/N} & \textbf{$\mathrm{GMRES}$} & \textbf{$\mathrm{FPSA_{GMRES}^{MPD}}$} & \textbf{$\mathrm{FPSA_{GMRES}^{WFD}}$} & \textbf{$\mathrm{FPSA_{SI}^{MPD}}$} & \textbf{$\mathrm{FPSA_{SI}^{WFD}}$} \\
\hline
GMRES & 15/16 & 2086 & 9 & 10 & 12 & 35 \\
					 & 15/32  & 1932 & 14 & 12 & 24 & 28 \\
Factorize & 15/16 & & 14 & 13 & 12 & 35 \\
					 &  15/32 & & 14 & 13 & 24 & 28 \\
\hline
\end{tabular}
\caption{EK - Beam Source - Number of Iterations}
\label{table:iter_comp_ek_b}
\end{table}
\linespread{2}
\linespread{1}
\begin{table}[H]
\centering
\begin{tabular}{c c c c c c c}
\hline \hline
\textbf{FP-Solve} & \textbf{L/N} & \textbf{$\mathrm{GMRES}$} & \textbf{$\mathrm{FPSA_{GMRES}^{MPD}}$} & \textbf{$\mathrm{FPSA_{GMRES}^{WFD}}$} & \textbf{$\mathrm{FPSA_{SI}^{MPD}}$} & \textbf{$\mathrm{FPSA_{SI}^{WFD}}$} \\
\hline
GMRES & 15/16 & 39.55 & 9.45 & 11.63 & 8.335 & 19.42 \\
					 & 15/32 & 36.14 & 24.89 & 18.57 & 34.74 & 24.87 \\
Factorize & 15/16 & & 2.842 & 6.657 & 0.2929 & 0.6853 \\
					  & 15/32  & & 2.898 & 6.799 & 0.6329 & 0.6585 \\
\hline
\end{tabular}
\caption{EK - Beam Source - Runtime [s]}
\label{table:rt_comp_ek_b}
\end{table}
\linespread{2}
We see similar behavior to what we saw in the case of SRK. The solver runtimes differ due to difference in rate at which FP-solve converges for this particular problem. Again, we note a significant decrease in number of iterations but a decrease in solver runtime strongly depends on the efficiency of the FP-solve.

\subsection{Henyey-Greenstein Kernel}

In this section, we let the asymmetry parameter, $g = 0.9999$. The study is carried out in the same way as the previously for SRK and EK. For this section, we will choose $\sigma_a = 0.00001$ $cm^{-1}$. The scattering cross-section moments are calculated using HGK. We will choose slab length of $50$ cm disretized using $200$ elements. Number of iterations and overall runtime data has been presented in Table \ref{table:iter_comp_hgk_v}, \ref{table:rt_comp_hgk_v}, \ref{table:iter_comp_hgk_b}, and \ref{table:rt_comp_hgk_b}.  \\
\linespread{1}
\begin{table}[H]
\centering
\begin{tabular}{c c c c c c c}
\hline \hline
\textbf{FP-Solve} & \textbf{L/N} & \textbf{$\mathrm{GMRES}$} & \textbf{$\mathrm{FPSA_{GMRES}^{MPD}}$} & \textbf{$\mathrm{FPSA_{GMRES}^{WFD}}$} & \textbf{$\mathrm{FPSA_{SI}^{MPD}}$} & \textbf{$\mathrm{FPSA_{SI}^{WFD}}$} \\
\hline
GMRES & 15/16 & 1150 & 10 & 7 & 28 & 18 \\
	  & 15/32  & 1461 & 14 & 12 & 32 & 31 \\
Factorize & 15/16 & & 10 & 8 & 28 & 18 \\
		  &  15/32 & & 14 & 13 & 32 & 31 \\
\hline
\end{tabular}
\caption{HGK - Vacuum Boundaries/Unit Distributed Source - Number of Iterations}
\label{table:iter_comp_hgk_v}
\end{table}
\linespread{2}
\linespread{1}
\begin{table}[H]
\centering
\begin{tabular}{c c c c c c c}
\hline \hline
\textbf{Invert FP} & \textbf{L/N} & \textbf{$\mathrm{GMRES}$} & \textbf{$\mathrm{FPSA_{GMRES}^{MPD}}$} & \textbf{$\mathrm{FPSA_{GMRES}^{WFD}}$} & \textbf{$\mathrm{FPSA_{SI}^{MPD}}$} & \textbf{$\mathrm{FPSA_{SI}^{WFD}}$} \\
\hline
GMRES & 15/16 & 83.66 & 288.1 & 160.9 & 931.9 & 373.4 \\
	  & 15/32  & 102.5 & 1055 & 572.7 & 2618 & 1428 \\
Factorize & 15/16 & & 6.390 & 12.27 & 2.193 & 1.385 \\
		  & 15/32  & & 11.75 & 28.820 & 2.651 & 2.5778 \\
\hline
\end{tabular}
\caption{HGK - Vacuum Boundaries/Unit Distributed Source - Runtime [s]}
\label{table:rt_comp_hgk_v}
\end{table}
\linespread{2}
%
% BEAM ===============================================================
%
\linespread{1}
\begin{table}[H]
\centering
\begin{tabular}{c c c c c c c}
\hline \hline
\textbf{Invert FP} & \textbf{L/N} & \textbf{$\mathrm{GMRES}$} & \textbf{$\mathrm{FPSA_{GMRES}^{MPD}}$} & \textbf{$\mathrm{FPSA_{GMRES}^{WFD}}$} & \textbf{$\mathrm{FPSA_{SI}^{MPD}}$} & \textbf{$\mathrm{FPSA_{SI}^{WFD}}$} \\
\hline
GMRES & 15/16 & 597 & 16 & 12 & 29 & 17 \\
					 & 15/32  & 1634 & 21 & 19 & 32 & 31 \\
Factorize & 15/16 & & 12 & 9 & 29 & 17 \\
					 &  15/32 & & 17 & 16 & 32 & 31 \\
\hline
\end{tabular}
\caption{HGK - Beam Source - Number of Iterations}
\label{table:iter_comp_hgk_b}
\end{table}
\linespread{2}
\linespread{1}
\begin{table}[H]
\centering
\begin{tabular}{c c c c c c c}
\hline \hline
\textbf{FP-Solve} & \textbf{L/N} & \textbf{$\mathrm{GMRES}$} & \textbf{$\mathrm{FPSA_{GMRES}^{MPD}}$} & \textbf{$\mathrm{FPSA_{GMRES}^{WFD}}$} & \textbf{$\mathrm{FPSA_{SI}^{MPD}}$} & \textbf{$\mathrm{FPSA_{SI}^{WFD}}$} \\
\hline
GMRES & 15/16 & 42.29 & 479.9 & 331.2& 1040 & 285.7 \\
					 & 15/32  & 115.1 & 1804 & 990.6 & 2408 & 1315 \\
Factorize & 15/16 & & 6.795 & 12.41 & 2.131 & 1.316 \\
					  & 15/32  & & 12.59 & 29.69 & 2.945 & 2.452 \\
\hline
\end{tabular}
\caption{HGK - Beam Source - Runtime [s]}
\label{table:rt_comp_hgk_b}
\end{table}
\linespread{2}
We note that, just like for SRK and EK, preconditioned schemes have significantly less iteration counts. However depending on how the Fokker-Planck error equation is solved, the preconditioning may or may not be effective with respect to runtime reduction. Solving the FP equation with GMRES renders FPSA scheme unviable, however use of factorization reduces to overall runtime significantly.

\section{Summary and Future Work}

We ran several numerical experiments and assessed the speed-ups in iteration count and solver runtime. We saw that preconditioning transport solve using FP resulted in reduction in iteration count by upto three orders (when compared to unpreconditioned GMRES solves). The overall runtime, however, depended completely on how efficiently the FP preconditioner was solved. Direct factorization resulted in a runtime reduction by upto two orders of magnitude. We observed that FP can be a very effective preconditioner for transport solves with highly forward-peaked scattering. However, we must develop an effective solver for FP-solve itself in order to make this an attractive preconditioning method. In future, we would like to determine how do we optimize FP-solve. We would also like to test FPSA's performance in energy dependent, multi-D settings. Moreover, we would also like to develop a nonlinear version of this method which would allow us to obtain a Fokker-Planck equation that is consistent with the relevant transport equation. 
%
% +++++++++++++++++++++++++++++++++++++++++++++++++++++++++++++++++++++++++++++
%
\vspace*{1em}
\begin{center}
\textbf{Acknowledgments}
\end{center}

This information has been co-authored by an employee or employees
of the Los Alamos National Security, LLC.~(LANS), operator of the
Los Alamos National Laboratory under Contract No.~DE-AC52-06NA25396 with the
U.S.~Department of Energy. \\

\setlength{\baselineskip}{12pt}

% =============================================================================
%
% APPENDIX AND BIBLIOGRAPHY
%
% +++++++++++++++++++++++++++++++++++++++++++++++++++++++++++++++++++++++++++++
%
% NO APPENDIX (commented out for future reference)
%
%\newpage
%\clearpage
%
%\section{APPENDIX}
%
% +++++++++++++++++++++++++++++++++++++++++++++++++++++++++++++++++++++++++++++
% BIBLOGRAPHY
% =============================================================================

%\newpage
%\clearpage
%
%% uncomment for submission of manuscript to NSE
%\pagestyle{empty}
%
%\bibliography{refs}
%
%% +++++++++++++++++++++++++++++++++++++++++++++++++++++++++++++++++++++++++++++
%%
%% Figures and Tables
%%
%% =============================================================================
%
%% uncomment for submission of manuscript to NSE
%\newpage
%\clearpage
%\listof{figure}{Figures}
%
%% NOTE: no tables so comment out to eliminate an empty list
%
%\newpage
%\clearpage
%\listof{table}{Tables}

% +++++++++++++++++++++++++++++++++++++++++++++++++++++++++++++++++++++++++++++
%
% END OF DOCUMENT
%
% =============================================================================

\end{document}